\documentclass[pra,twocolumn,showpacs,floatfix,nofootinbib]{revtex4}

\usepackage{graphicx}
\usepackage{amsmath}
\usepackage{amsfonts}

\providecommand{\U}[1]{\protect\rule{.1in}{.1in}}

\newenvironment{proof}[1][Proof]{\noindent\textbf{#1.} }{\ \rule{0.5em}{0.5em}}

\newcommand{\be}{\begin{equation}}
\newcommand{\ee}{\end{equation}}

\newtheorem{mylemma}{Lemma}

\begin{document}

\title{Rigorous Bounds on the Performance of a Hybrid Dynamical
  Decoupling-Quantum Computing Scheme}
\author{Kaveh Khodjasteh}
\affiliation{Department of Physics, Center for Quantum Information Science \& Technology,
University of Southern California, Los Angeles, CA 90089\\
Department of Physics and Astronomy, Dartmouth College, Hanover, NH 03755}
\author{Daniel A. Lidar}
\affiliation{Departments of Chemistry, Electrical Engineering, and Physics, Center for
Quantum Information Science \& Technology, University of Southern
California, Los Angeles, CA 90089}

\begin{abstract}
We study dynamical decoupling in a multi-qubit setting, where it is combined
with quantum logic gates. This is illustrated in terms of computation using
Heisenberg interactions only, where global decoupling pulses commute with
the computation. We derive a rigorous error bound on the trace distance or
fidelity between the desired computational state and the actual time-evolved
state, for a system subject to coupling to a bounded-strength bath. The
bound is expressed in terms of the operator norm of the effective
Hamiltonian generating the evolution in the presence of decoupling and logic
operations. We apply the bound to the case of periodic pulse sequences and
find that in order maintain a constant trace distance or fidelity, the
number of cycles -- at fixed pulse interval and width -- scales in inverse
proportion to the square of the number of qubits. This sets a scalability
limit on the protection of quantum computation using periodic dynamical
decoupling.
\end{abstract}

\maketitle

\section{Introduction}

Quantum information processing harbors enormous unleashed potential in
the form of efficient algorithms for classically intractable tasks \cite%
{Nielsen:book}. Perhaps the largest hurdle on the way to a realization of
this potential is the problem of decoherence, which results when a quantum
system, such as a quantum computer, interacts with an uncontrollable
environment \cite{Breuer:book}. Decoherence reduces the information
processing capabilities of quantum computers to the point where they can be
efficiently simulated on a classical computer \cite{Aharonov:96a}. In spite
of dramatic progress in the form of a theory of fault tolerant quantum error
correction (e.g., \cite{Aliferis:05}), finding methods for overcoming
decoherence that are both efficient and practical remains an important
challenge. An alternative to quantum error correction (QEC)\ that is
substantially less resource-intensive is dynamical decoupling (DD) \cite%
{Viola:98}. This method does not require feedback or the exponential growth
in the number of qubits typical of fault tolerant (concatenated) QEC. In DD
one applies a succession of short pulses to the system, designed to decouple
it from the environment. This can substantially slow down decoherence,
though not halt it completely, in contrast to the promises of fault-tolerant
QEC. While initially the general theory of DD\ was developed under the
assumption of highly idealized (essentially infinitely fast and strong)
pulses \cite{Zanardi:98b,Viola:99,Facchi:03}, subsequent work relaxed these
assumptions, showing that DD\ can still be beneficial in the presence of
bounded strength controls \cite{Viola:02}. In the simplest possible DD\
protocol, known as \textquotedblleft periodic DD\textquotedblright\ (PDD),
one applies a certain predetermined sequence over and over again. While this
protocol typically does not work as well as random \cite%
{Viola:05,Viola:06,Santos:06}, recursive-deterministic \cite{KL1}, or hybrid
schemes \cite{kern:250501} when finite pulse intervals and pulse width are
accounted for, it has the advantage of simplicity. In this work our purpose
is to present a rigorous analysis of DD, and in particular to derive error
bounds on its performance in the periodic (PDD) setting.

With a few exceptions that belong to the realm of idealized pulses \cite%
{Viola:99a,ByrdLidar:01a,Viola:01a,LidarWu:02} or to the paradigm of
adiabatic quantum computing \cite{Lidar:AQC-DD}, DD\ studies have focused on
preserving quantum information (memory), rather than processing it
(computation). 
In order to combine computation with DD, Ref. \cite{Viola:99a}
introduced three strategies: The first strategy requires applying the
computational operations \textquotedblleft stroboscopically\textquotedblright , i.e., at the end of each decoupling
cycle, where the system is momentarily decoherence-free. This is
conceptually similar to computation over error-correcting codes, where a
computational gate is applied at the end of an error-correction cycle \cite%
{Gottesman:97a}. The disadvantage of this \textquotedblleft
stroboscopic\textquotedblright\ approach is that, in reality, the
computational operations take a finite time to implement, so that the system
decoheres while a computational gate is being applied to it. 
The second strategy, is to alternate and modulate the control
Hamiltonian used to implement quantum computation, in which the net
overall effect of the DD operations still allows a desired unitary
operation on the system, along with the correction of
errors. The third strategy proposed in Ref. \cite{Viola:99a} is to use DD pulses that commute with the
computational operations, so that the two can be executed simultaneously.
Here we address the problem of circuit-model quantum computation \cite%
{Nielsen:book} using DD with realistic pulse assumptions. We combine DD with
computation via the use of codes and universality results arising from the
theory of decoherence-free subspaces and subsystems (DFSs) \cite%
{Zanardi:97c,Knill:99a,Kempe:00}. Our method is conceptually related
to a hybrid version of the second and third strategies of Ref.~\cite{Viola:99a}, in that we impose the
commutation condition between the DD pulses and the computational
Hamiltonian, but we find that improved performance is obtained if the DD and
computational operations simply alternate. Thus, in our scheme the computational
gate is \textquotedblleft spread\textquotedblright\ over an entire DD\
cycle (or, conversely, the DD\ cycle is spread over the computational gate).
We fully incorporate finite pulse intervals and pulse widths and assess the
performance of our scheme in the PDD setting. We find a rigorous error
bound, from which it follows that for a fixed error the number of DD\ cycles
cannot scale faster than the inverse square of the system size (at fixed
pulse width and pulse interval). This means that there is a tradeoff between
the length of time over which decoherence errors can be suppressed using
PDD, and the scalability of a quantum computation it is meant to protect.

The structure of this paper is as follows. We define the model in Section %
\ref{sec:model}. We provide background on dynamical decoupling in Section %
\ref{sec:DD-back}, where we also derive the effective Hamiltonian describing
the evolution under the action of decoupling and computation. This leads to
the \textquotedblleft error phase\textquotedblright , namely the effective
Hamiltonian times time (a type of action), which is the quantity we wish to
minimize. In Section \ref{sec:bounds} we derive rigorous error bounds that
relate the error between the desired and actual final state to the norm of
the error phase. In Section \ref{sec:error} we estimate the error associated
with the decoupled evolution (i.e., the evolution in the presence of a DD\
pulse sequence), relative to the decoherence-free evolution (no system-bath
coupling). Section \ref{sec:long} is where we derive our key result:\ we
apply the idea of encoded operations and dynamical decoupling to PDD, and
compute the error bound. In Section \ref{sec:Heis-DD} we illustrate our
construction with encoded DD-computation in a quantum dots setting, where
computation is implemented via Heisenberg interactions. We conclude with a
discussion of our results in Section \ref{sec:conc}. Extensive background
material is presented in the appendices.

\section{Model}

\label{sec:model}

We express the total Hamiltonian for system plus bath in the form%
\begin{equation}
H(t)=H_{\text{ctrl}}(t)\otimes I_{B}+H_{\text{err}}+I_{S}\otimes H_{B}
\label{eq:hdogma}
\end{equation}
where $I$ is the identity operator, $H_{\text{ctrl}}$ acts on the system
only and serves to implement (encoded) control operations such as logic
gates, $H_{\text{err}}$ is the {}``error\textquotedblright\ Hamiltonian
(system-bath couplings, undesired interactions among system qubits that do
not commute with $H_{\text{ctrl}}$), and $H_{B}$ is the pure-bath
Hamiltonian. Let $U_{\text{ctrl}}$ be the (encoded) logic gate generated by
switching on $H_{\text{ctrl}}$ for duration $T$, in the absence of the bath
and any undesired interactions within the system:%
\begin{equation}
U_{\text{ctrl}}(T)=\mathcal{T}\exp\left[-i\int_{0}^{T}H_{\text{ctrl}}(t)dt%
\right]=\exp(-i\theta R),  \label{eq:encodedR}
\end{equation}
where $\mathcal{T}$ denotes time-ordering, $R$ is a dimensionless logic
operator, and $\theta$ is the angle of rotation around this operator.
However, due to the presence of the undesired $H_{\text{err}}$ and $H_{B}$
terms, we will in fact obtain the following unitary acting on the joint
system and bath Hilbert space:%
\begin{equation}
U_{\text{bare}}(T)=\mathcal{T}\exp\left[-i\int_{0}^{T}\left(H_{\text{ctrl}%
}(t)+H_{\text{err}}+H_{B}\right)dt\right],  \label{eq:bareU}
\end{equation}
This is the essence of the problem of any quantum control procedure, whether
it be for quantum information processing or other purposes:\ $U_{\text{bare}%
} $ entangles system and bath and implements a tranformation on the system
that can be very different from the desired $U_{\text{ctrl}}$. Our goal in
this work is to show how to modify $U_{\text{bare}}$ so that the distance
between a state evolving under it and a state evolving under $U_{\text{ctrl}%
} $ can be made arbitrarily small. This will be done by adding another
Hamiltonian to the system, which implement DD\ operations, and is designed
to effectively cancel $H_{\text{err}}$ without interfering with $H_{\text{%
ctrl}}$.

\section{Dynamical Decoupling Background and the Error Phase}

\label{sec:DD-back}

\subsection{Dynamical Decoupling Defined}

We assume that the decoupling operations are realized as pulses $P_{i}$ by a
time-dependent Hamiltonian $H_{\text{DD}}(t)$. The essential condition that
will ensure that the decoupling pulses interfere minimally with the control
operations is:%
\begin{equation}
\lbrack H_{\text{DD}}(t),H_{\text{ctrl}}(t^{\prime })]=0\quad \forall
t,t^{\prime }.  \label{eq:commcond}
\end{equation}%
The total propagator is now generated by the time-dependent total Hamiltonian%
\begin{equation}
H_{\mathrm{tot}}(t)=H_{\text{DD}}(t)+H_{\text{ctrl}}(t)+H_{\text{err}}+H_{B},
\label{eq:H(t)}
\end{equation}%
i.e.,\
\begin{equation}
U(T)=\mathcal{T}\exp \left[ -i\int_{0}^{T}H_{\mathrm{tot}}(t)dt\right] .
\label{eq:U}
\end{equation}%
The pulses are applied at times $\left\{ t_{j}\right\} _{j=0}^{N}$ given by%
\begin{equation}
t_{j}=j(\tau +\delta ),
\end{equation}%
where $\tau $ is the pulse interval and $\delta $ is the pulse width. From
hereon we assume for simplicity that $H_{\text{DD}}(t)$ is piecewise
constant (thus we are performing a worst-case analysis: our conclusions can
only be improved by pulse shaping \cite{Kofman:01,sengupta:037202}):%
\begin{eqnarray}
H_{\text{DD}}(t) &=&%
\begin{cases}
0 & t_{j}\leq t<t_{j+1}-\delta \\
H_{P}^{(j+1)} & t_{j+1}-\delta \leq t<t_{j+1}%
\end{cases}
\label{eq:Hddoft} \\
j &\in &\{0,\cdots ,N-1\}.  \notag
\end{eqnarray}%
Let
\begin{equation}
T=t_{N}=N(\tau +\delta )
\end{equation}
denote the time it takes to complete one DD cycle, consisting of $N$ pulses
generated by the $H_P^j$:
\begin{equation}
P_{j}=\exp (-iH_{P}^{(j)}\delta )\quad j=1,...,N.  \label{eq:Pj}
\end{equation}%
The commutation condition (\ref{eq:commcond}) becomes%
\begin{equation}
\lbrack H_{P}^{(j)},H_{\text{ctrl}}(t)]=[P_{j},H_{\text{ctrl}}(t)]=0 \quad
\forall j.
\end{equation}%
This will allow us to import many of the results of the control-free
scenario, i.e., when $H_{\text{ctrl}}(t)=0$. For the remainder of this
section we review this setting, and return to the question of how to ensure
the commutation condition in section \ref{sec:Heis-DD}.

Denoting a free evolution period (when $H_{\text{DD}}=0$) of duration $\tau $
by
\begin{equation}
\mathtt{f}_{\tau }=\exp \left[ -i\tau \left( H_{\text{err}}+H_{B}\right) %
\right] ,  \label{eq:f_tau}
\end{equation}%
a single cycle can be written as%
\begin{widetext}
\begin{eqnarray}
P_{N}\mathtt{f}_{\tau }P_{N-1}\mathtt{f}_{\tau }P_{N-2}\cdots P_{1}\mathtt{%
f}_{\tau }
&=&P_{N}\mathtt{f}_{\tau }[P_{N}^{\dag }P_{N}]P_{N-1}\mathtt{f}_{\tau
}[(P_{N}P_{N-1})^{\dagger }(P_{N}P_{N-1})]P_{N-2}\cdots P_{1}\mathtt{f}%
_{\tau }  \notag \\
&=&(D_{N}\mathtt{f}_{\tau }D_{N}^{\dag })(D_{N-1}\mathtt{f}_{\tau
}D_{N-1}^{\dag })D_{N-2}\cdots D_{1}\mathtt{f}_{\tau }D_{1}^{\dag }
\notag \\
&\equiv&
e^{-iTH_{\mathrm{eff}}^{(1)}(T)},
\label{eq:genericseq}
\end{eqnarray}%
\end{widetext}
where the unitary {}\textquotedblleft decoupling group\textquotedblright\
\cite{Zanardi:98b} $\mathcal{G}=\{D_{j}\}_{j=1}^{N}$ has elements defined as

\begin{equation}
D_{j}\equiv P_{N}\cdots P_{j},\quad D_{1}\equiv I_{S},  \label{eq:D}
\end{equation}%
where the condition $D_{1}\equiv I_{S}$ is imposed because of the appearance
of $D_{1}^{\dag }$ in Eq.~(\ref{eq:genericseq}); this imposes a relation
among the pulse Hamiltonians $H_{P}^{(j)}$ via Eq.~(\ref{eq:Pj}). Note that
this is only possible in the zero width limit, since such a relation cannot
be satisfied when the system-bath and bath Hamiltonians are present during
the pulse.

The effective Hamiltonian $H_{\mathrm{eff}}^{(1)}(T)$ can be
computed using the Magnus expansion \cite{Iserles:02} (see also Appendix \ref%
{app:Magnus}). To first order in the Magnus expansion
\begin{eqnarray}
H_{\mathrm{eff}}^{(1)}(T) &\equiv &\sum_{j=1}^{N}D_{j}(H_{%
\text{err}}+H_{B})D_{j}^{\dag }  \notag \\
&=&\Pi _{\mathcal{G}}(H_{\text{err}})+NH_{B}.
\end{eqnarray}%
This can be viewed as a projection
\begin{equation}
\Pi _{\mathcal{G}}(H_{\text{err}})\equiv \sum_{j=1}^{N}D_{j}H_{\text{err}%
}D_{j}^{\dag }  \label{eq:proj}
\end{equation}%
into the centralizer $Z(\mathcal{G})\equiv \{V|\;[V,D_{j}]=0,\;\forall
D_{j}\in \mathcal{G}\}$. Indeed, since $[H_{B},D_{j}]=0$ for all $j$,
\begin{eqnarray}
D_{k}^{\dag }[H_{\mathrm{eff}}^{(1)},D_{k}] &=&D_{k}^{\dag
}\sum_{j=1}^{N}[D_{j}H_{\text{err}}D_{j}^{\dag }+H_{B},D_{k}]  \notag \\
&=&\sum_{j=1}^{N}(D_{k}^{\dag }D_{j})H_{\text{err}}(D_{k}^{\dag
}D_{j})^{\dag }-D_{j}H_{\text{err}}D_{j}^{\dag }  \notag \\
&=&0  \notag \\
&\Longrightarrow &[H_{\mathrm{eff}}^{(1)},D_{k}]=0\quad \forall k,
\end{eqnarray}%
where in the last equality we used the group closure property: $\forall k,j$
$\exists i$ such that $D_{k}^{\dag }D_{j}=D_{i}$. For a unitary irreducible
representation of $\mathcal{G}$ this immediately implies, by Schur's lemma,
that $H_{\mathrm{eff}}^{(1)}=cI_{S}\otimes H_{B}$ where $c$ is a constant,
i.e., $H_{\mathrm{eff}}^{(1)}$ acts harmlessly on the system.

If the algebraic structure of  $H_{\text{err}}$ is known, we can choose the decoupling group such that it
satisfies the {}\textquotedblleft decoupling condition\textquotedblright\ for any Hamiltonian $H_{\text{err}}$:
\cite{Zanardi:98b,Viola:99}
\begin{equation}
\Pi _{\mathcal{G}}(H_{\text{err}}) = \sum_{j=1}^{N}D_{j}H_{\text{err}%
}D_{j}^{\dag }=0.  \label{eq:DDcond}
\end{equation}%
In the limit $\tau ,\delta \rightarrow 0$, and in the absence of control,
the first order Magnus expansion is exact and condition (\ref{eq:DDcond})
guarantees the stroboscopic elimination of $H_{\text{err}}$, in the sense
that $H_{\mathrm{eff}}^{(1)}(T)=0$, and this would be true at
the end of every DD\ cycle. Another way to understand condition (\ref%
{eq:DDcond}) is to recall that $D_{1}=I_{S}$, which means that $%
\sum_{j=2}^{N}D_{j}H_{\text{err}}D_{j}^{\dag }=-H_{\text{err}}$: the
negative sign in front of $H_{\text{err}}$ means that the role of the
decoupling group is to effectively time-reverse the error Hamiltonian at the
end of the cycle.

\subsection{Interaction Picture}

\label{IP}

In a setting where decoupling works perfectly the system evolves
independently from the bath, purely under the action of the control
Hamiltonian. Therefore we use the interaction picture of
\begin{equation}
H_{\text{sec}}\equiv H_{\text{ctrl}}+H_{B}
\end{equation}
(the sum of the secular terms) to calculate the full propagator {[}Eq.~(\ref%
{eq:U})]:%
\begin{equation}
U(t)=U_{\text{sec}}U_{\text{err}}(t,0),
\end{equation}%
where%
\begin{eqnarray}
U_{\text{sec}} &\equiv &U_{\text{ctrl}}(t)\otimes U_{B}(t) \\
U_{\text{ctrl}}(t) &=&\mathcal{T}\exp \left[ -i\int_{0}^{t}H_{\text{ctrl}%
}(s)ds\right] , \\
U_{B}(t) &=&\exp (-itH_{B}).
\end{eqnarray}%
If $|\Psi (t)\rangle =U(t)|\Psi (0)\rangle $ (Schr\"{o}dinger picture)\ then
$|\tilde{\Psi}(t)\rangle =U_{\text{sec}}^{\dag }|\Psi (t)\rangle =U_{\text{%
err}}(t)|\Psi (0)\rangle $ is the corresponding state in the interaction
picture. Similarly, for mixed states:\ $\tilde{\rho}(t)=U_{\text{sec}}^{\dag
}\rho (t)U_{\text{sec}}=U_{\text{err}}(t,0)\rho (0)U_{\text{err}}^{\dag
}(t,0)$. The interaction picture propagator $U_{\text{err}}$ contains all
the {}\textquotedblleft errors\textquotedblright , in the sense that if it
becomes the identity operator then decoupling is perfect. For then $U(t)=U_{%
\text{ctrl}}(t)\otimes U_{B}(t)$ and the desired system dynamics is
completely decoupled from the bath. In this sense the interaction picture is
naturally suited to our analysis:\ by moving the {}\textquotedblleft
ideal\textquotedblright\ evolution $U_{\text{ctrl}}(t)\otimes U_{B}(t)$ to
the left, we have isolated the {}\textquotedblleft error
propagator\textquotedblright\ $U_{\text{err}}$.

$U_{\text{err}}$ satisfies the Schr\"{o}dinger equation
\begin{equation}
\frac{dU_{\text{err}}(t,0)}{dt}=-i\tilde{H}_{\text{err}}(t)U_{\text{err}%
}(t,0),\quad U_{\text{err}}(0,0)=I,  \label{eq:Uerr}
\end{equation}%
with
\begin{eqnarray}
\tilde{H}_{\text{err}}(t) &=&U_{B}^{\dag }U_{\text{ctrl}}^{\dag }[H_{\mathrm{%
DD}}(t)+H_{\text{err}}]U_{B}U_{\text{ctrl}}  \notag \\
&=&H_{\mathrm{DD}}(t)+\text{Ad}_{tH_{\text{sec}}}[H_{\text{err}}],
\label{eq:Htilde-err}
\end{eqnarray}%
where the linear adjoint map $\text{Ad}_{A}[B]$ has the
Baker-Campbell-Hausdorff formula\cite{Reinsch:00}%
\begin{equation}
\text{Ad}_{A}[B]\equiv e^{-iA}Be^{iA}=\sum_{n=0}^{\infty }\frac{(-i)^{n}}{n!}%
[_{n}A,B],  \label{eq:adab}
\end{equation}%
where $[_{n}A,B]$ denotes a nested commutator term $[A,[\cdots \lbrack A,B]$
in which $A$ appears $n$ times. Note that -- thanks to the commutation
condition (\ref{eq:commcond}) -- $H_{\mathrm{DD}}(t)$ remains invariant
under the interaction picture transformation in Eq.~(\ref{eq:Htilde-err}).
This is where the commutation condition shows up explicitly in our analysis.

Let us define an effective (dimensionless) {}``error
phase\textquotedblright\ $\Phi_{E}$ via \cite{KhodjastehLidar:07}:
\begin{equation}
\exp[-i\Phi_{E}(T)]\equiv U_{\text{err}}(T,0).  \label{eq:Phi_E}
\end{equation}
Thus $\Phi_{E}(T)$ is the final effective Hamiltonian times the total time,
and it measures the deviation from ideal dynamics. In other words, the goal
of the decoupling procedure is to minimize $\Phi_{E}(T)$. In Section \ref%
{sec:bounds} we relate $\Phi_{E}$ to conventional fidelity measures.
Throughout this work we repeatedly use the technique of expressing unitaries
in terms of the {}``final effective Hamiltonian\textquotedblright . In fact,
this was already done in our review of DD\ above, when we used the effective
Hamiltonian $H_{\mathrm{eff}}^{(1)}(T)$ in Eq.~(\ref%
{eq:genericseq}).

\subsection{The Error Phase}

We now wish to calculate the total propagator $U$ {[}Eq.~(\ref{eq:U})] in
the presence of both decoupling and control. The evolution generated by $%
\tilde{H}_{\text{err}}(t)$ {[}Eq.~(\ref{eq:Htilde-err})] can be decomposed
into {}{}\textquotedblleft free\textquotedblright\ and pulse periods as
follows:
\begin{eqnarray}
&&\tilde{H}_{\text{err}}(t)  \label{eq:Herrtilde} \\
&=&%
\begin{cases}
\text{Ad}_{tH_{\text{sec}}}(H_{\text{err}}) & \text{ }t_{i-1}<t<t_{i-1}+\tau
\\
H_{P}^{(i)}+\text{Ad}_{tH_{\text{sec}}}(H_{\text{err}}) & \text{ }%
t_{i-1}+\tau <t<t_{i} \notag%
\end{cases}%
.
\end{eqnarray}%
In Appendix \ref{app:Udecomp} we prove the following \textquotedblleft
switching lemma\textquotedblright :

\begin{mylemma}
\label{lem1}

The propagator generated by a ``switched Hamiltonian\textquotedblright
\begin{equation}
H(t)=H_{i}(t)\quad t_{i-1}<t<t_{i},\quad i=1,...,N,
\end{equation}
can be decomposed into corresponding segments:%
\begin{equation}
U(t_{N},t_{0})=U(t_{N},t_{N-1})\cdots U(t_{1},t_{0}),  \label{eq:decomp}
\end{equation}
where $U(t_{i+1},t_{i})$, with $t_{i}\leq t\leq t_{i+1}$, satisfies
the Schr\"{o}dinger equation 
\begin{equation}
\frac{dU(t,t_{i})}{dt}=-iH(t)U(t,t_{i}),\quad U(t_{i},t_{i})=I.
\label{eq:Schr}
\end{equation}
\end{mylemma}

We can thus write:%
\begin{equation}
U_{\text{err}}(T,0)=U_{\text{err}}(t_{N},t_{N-1})\cdots U_{\text{err}%
}(t_{1},t_{0}),  \label{eq:Udecomp}
\end{equation}%
where $U_{\text{err}}(t_{i},t_{i-1})$, with $t_{i-1}\leq t\leq t_{i}$,
satisfies the Schr\"{o}dinger equation
\begin{eqnarray}
\frac{dU_{\text{err}}(t,t_{i-1})}{dt} &=&-i\tilde{H}_{\text{err}}(t)U_{\text{%
err}}(t,t_{i-1})\quad  \label{eq:SchrUi} \\
U_{\text{err}}(t_{i-1},t_{i-1}) &=&I,  \notag
\end{eqnarray}%
whose formal solution is%
\begin{equation}
U_{\text{err}}(t,t_{i-1})=\mathcal{T}\exp [-i\int_{t_{i-1}}^{t}\tilde{H}_{%
\text{err}}(t)dt].
\end{equation}

Analogously, we can further decompose each segment into a pulse and a free
evolution, each with an effective Hamiltonian. The pulse part is
\begin{eqnarray}
U_{\text{err}}(t_{i},t_{i-1}) &=&U_{\text{err}}(t_{i},t_{i}-\delta )U_{\text{%
err}}(t_{i}-\delta ,t_{i-1})  \notag \\
&=&\mathcal{T}e^{-i\int_{t_{i}-\delta }^{t_{i}}\tilde{H}_{\text{err}%
}(t)dt}U_{\text{err}}(t_{i}-\delta ,t_{i-1})  \notag \\
&=&P_{i}e^{-i\delta H_{\text{err}}^{P_{i}}}U_{\text{err}}(t_{i}-\delta
,t_{i-1}),
\end{eqnarray}%
which, using Eq.~(\ref{eq:Herrtilde}), serves to define the effective error
\begin{equation}
\exp [-i\delta H_{\text{err}}^{P_{i}}(\delta )]\equiv P_{i}^{\dag }\mathcal{T%
}e^{-i\int_{t_{i}-\delta }^{t_{i}}(H_{P}^{(i)}+\text{Ad}_{tH_{\text{sec}%
}}(H_{\text{err}}))dt}  \label{eq:H_err^Pi}
\end{equation}%
associated with the width of the ideal pulse $P_{i}=\exp (-i\delta
H_{P}^{(i)})$.

The $i$-th free segment $(t_{i-1},t_{i}-\delta )$ is similarly generated by
an effective Hamiltonian $H_{\text{err}}^{(i)}$ defined via
\begin{eqnarray}
\exp [-i\tau H_{\text{err}}^{(i)}] &\equiv &U_{\text{err}}(t_{i-1}+\tau
,t_{i-1})  \notag \\
&=&\mathcal{T}e^{-i\int_{0}^{\tau }\text{Ad}_{(s+t_{i-1})H_{\text{sec}}}[H_{%
\text{err}}]ds}.  \label{eq:H_err^i}
\end{eqnarray}%
The overall error unitary $U_{\text{err}}(T,0)$ can thus be written as%
\begin{eqnarray}
U_{\text{err}}(T,0) &=&P_{N}\exp [-i\delta H_{\text{err}}^{P_{N}}]\exp
[-i\tau H_{\text{err}}^{(N)}]  \notag \\
\cdots &\times &P_{1}\exp [-i\delta H_{\text{err}}^{P_{1}}]\exp [-i\tau H_{%
\text{err}}^{(1)}]  \label{eq:decU}
\end{eqnarray}%
To incorporate the effect of the DD operations we recall the definition of
the decoupling group in terms of the pulse unitaries {[}Eq.~(\ref{eq:D})]
and rewrite Eq.(\ref{eq:decU}) as

\begin{eqnarray}
U_{\text{err}}(T,0) &=&D_{N}e^{-i\delta H_{\text{err}}^{P_{N}}}D_{N}^{%
\dagger }D_{N}e^{-i\tau H_{\text{err}}^{(N)}}D_{N}^{\dagger }  \notag \\
\cdots &\times &D_{1}e^{-i\delta H_{\text{err}}^{P_{1}}}D_{1}^{\dagger
}D_{1}e^{-i\tau H_{\text{err}}^{(1)}}D_{1}^{\dagger }  \notag \\
&=&\prod_{j=1}^{N}\exp (-i\delta D_{j}^{\dagger }H_{\text{err}%
}^{P_{N}}D_{j}).  \label{eq:Uerr2}
\end{eqnarray}%
By Lemma~\ref{lem1} , the following
time-dependent Hamiltonian generates $U_{\text{err}}(T,0)$:%
\begin{equation}
H_{\text{m}}(t)\equiv
\begin{cases}
D_{i}H_{\text{err}}^{(i)}D_{i}^{\dagger } & \text{ for }t_{i}<t<t_{i}+\tau
\\
D_{i}H_{\text{err}}^{P_{i}}D_{i}^{\dagger } & \text{ for }t_{i}+\tau
<t<t_{i+1}%
\end{cases}%
.  \label{eq:Hm}
\end{equation}%
The {}\textquotedblleft free evolution error-Hamiltonian\textquotedblright\ $%
H_{\text{err}}^{(i)}$ and {}\textquotedblleft pulse
error-Hamiltonian\textquotedblright\ $H_{\text{err}}^{P_{i}}$ are defined,
respectively, in Eqs. (\ref{eq:H_err^i}) and (\ref{eq:H_err^Pi}). Gathering
our results we can write:

\begin{eqnarray}
\exp [-i\Phi _{E}(T)] &=&U_{\text{err}}(T,0)  \notag \\
&=&\mathcal{T}\exp [-i\int_{0}^{T}H_{\text{m}}(t)dt].  \label{eq:Uerr-final}
\end{eqnarray}

\section{Error Bounds}

\label{sec:bounds}

In this section we derive rigorous error bounds that relate the error
between the desired and actual final state to the norm of the error phase $%
\Phi _{E}(T)$. Throughout this work we use the trace distance
\begin{equation}
D(\rho _{1},\rho _{2})\equiv \frac{1}{2}\|\rho _{1}-\rho _{2}\|_{1},
\end{equation}
where
\begin{equation}
\|A\|_{1}\equiv \mathrm{tr}(\sqrt{A^{\dag }A}),
\end{equation}
as the distance measure between state, and the quantum fidelity, defined for
any pair of positive operators $A$ and $B$:%
\begin{equation}
F_{Q}(A,B)\equiv \|\sqrt{A}\sqrt{B}\|_{1}\overset{A^{\dag }=A,\,B^{\dag }=B}{%
=}\mathrm{tr}\sqrt{\sqrt{B}A\sqrt{B}}.
\end{equation}%
There is a useful relation between the trace distance and the quantum
fidelity \cite{Nielsen:book}:%
\begin{equation}
1-D(\rho _{1},\rho _{2})\leq F_{Q}(\rho _{1},\rho _{2})\leq \sqrt{1-D(\rho
_{1},\rho _{2})^{2}},  \label{eq:Qfid}
\end{equation}%
which means that the trace distance and fidelity can be used to bound one
another from below and above.

When one or more of the states is pure ($|\psi _{1}\rangle $, $|\psi
_{2}\rangle $), we shall write $D(\psi _{1},\psi _{2})$ and $F(\psi
_{1},\psi _{2})$, or use a mixed notation $D(\psi _{1},\rho _{2})$ and $%
F(\psi _{1},\rho _{2})$, etc.

We also make repeated use of the operator norm
\begin{equation}
\|A\|_{\infty }\equiv \sup_{\|\psi \|=1}\|A|\psi \rangle \|.
\end{equation}
For a review of these measures along with key properties see Appendix \ref%
{app:norms}.

In the absence of the bath the control Hamiltonian $H_{\text{ctrl}}(t)$
would implement a quantum computation via the propagator $U_{\text{ctrl}}$ {[%
}Eq.~(\ref{eq:encodedR})]. Equivalently, the state of the quantum computer
at the final time $T$ would be described by the solution $|\psi (T)\rangle $
of the Schr\"{o}dinger equation $|\dot{\psi}\rangle =-iH_{\mathrm{ctrl}%
}|\psi \rangle $. Imperfect control of $H_{\text{ctrl}}(t)$ means that even
in the absence of the bath, $|\psi (T)\rangle $ is not the ideal final
state, which would be obtained if one could implement a completely accurate
and noise-free Hamiltonian $H_{\text{ctrl}}^{\mathrm{ideal}}(t)$, with
corresponding final state $|\phi (T)\rangle $. Minimization of the
corresponding closed-system control error
\begin{equation}
\delta _{\mathrm{id}}\equiv D[\psi (T),\phi (T)]  \label{eq:del-id}
\end{equation}%
belongs to the realm of fault-tolerant quantum computation \cite{Preskill:99}
and composite pulse techniques \cite{Brown:04}, and will not be addressed
here.

The initial bath state is $\rho _{B}(0)$ and in the absence of coupling to
the system it evolves under the pure-bath Hamiltonian to $\rho
_{B}^{0}(t;\theta )=U_{B}(t)\rho _{B}^{0}(\theta )U_{B}(t)^{\dag }$ (the
superscript $0$ denotes no system-bath coupling).

In the general mixed-state setting we distinguish between
{}\textquotedblleft ideal\textquotedblright\ system evolution described by a
pure state $\rho _{S}^{\mathrm{ideal}}(t)=|\phi (t)\rangle \langle \phi (t)|$%
, with $|\phi (t)\rangle =U_{\text{ctrl}}^{\mathrm{ideal}}(t)|\phi
(0)\rangle $ and $U_{\text{ctrl}}^{\mathrm{ideal}}$ generated by $H_{\text{%
ctrl}}^{\mathrm{ideal}}$, and bath-free non-ideal system evolution (due to
control errors), described by a mixed state $\rho _{S}^{0}(t)$ (the mixed
nature can be due to, e.g., the need to average over stochastic realizations
of unitary evolutions). In the absence of any coupling between system and
bath the joint initial states $\rho _{S}^{\mathrm{ideal}}(0)\otimes \rho
_{B}^{0}(0)$ or $\rho _{S}^{0}(0)\otimes \rho _{B}^{0}(0)$ evolve in the two
scenarios to $\rho ^{\mathrm{ideal}}(t)\equiv \rho _{S}^{\mathrm{ideal}%
}(t)\otimes \rho _{B}^{0}(t)$ or $\rho ^{0}(t)\equiv \rho _{S}^{0}(t)\otimes
\rho _{B}^{0}(t)$, respectively.

Then the final error due to imperfect control in the uncoupled setting is%
\begin{equation}
D_{\mathrm{id}}\equiv D[\rho ^{0}(T),\rho ^{\mathrm{ideal}}(T)]=D[\rho
_{S}^{0}(T),\rho _{S}^{\mathrm{ideal}}(T)],  \label{eq:Del-id}
\end{equation}%
where we have used the multiplicativity property (\ref{eq:mult}) and unitary
invariance. Minimization of the pure-system control error $D_{\mathrm{id}}$ {%
[}generalization of Eq.~(\ref{eq:del-id})] once again lies in the domain of
fault tolerant quantum error correction \cite{Preskill:99} and composite
pulse techniques \cite{Brown:04}.

The actual system-bath state obtained by time evolution under the full
propagator $U(t)$ {[}Eq.~(\ref{eq:U})] is $\rho (t)=U(t)[\rho _{S}(0)\otimes
\rho _{B}(0)]U(t)^{\dag }$, and the actual system state is $\rho _{S}(t)=%
\mathrm{tr}_{B}\rho (t)$. \emph{The distance we wish to minimize is the
distance between the actual final system state} $\rho (t)$\emph{\ and the
ideal final system state} (no control errors, no coupling to the bath):%
\begin{equation}
D_{S}\equiv D[\rho _{S}(T),\rho _{S}^{\mathrm{ideal}}(T)].
\end{equation}%
Define
\begin{equation}
D_{\mathrm{tot}}\equiv D[\rho (T),\rho ^{\mathrm{ideal}}(T)].
\label{eq:Dtot}
\end{equation}%
By virtue of Eq.~(\ref{eq:part-trace}) we know that removing the partial
trace can only increase the distance between states, i.e.,%
\begin{equation}
D_{S}\leq D_{\mathrm{tot}}.
\end{equation}%
Let
\begin{equation}
D_{\mathrm{DD}}\equiv D[\rho (T),\rho ^{0}(T)]=D[\tilde{\rho}(T),\tilde{\rho}%
^{0}(0)],
\end{equation}%
where we have used the fact that in the interaction picture $\tilde{\rho}%
^{0}(t)=\tilde{\rho}^{0}(0)=\rho _{S}(0)\otimes \rho _{B}(0)$. $D_{\mathrm{DD%
}}$ is the distance due to coupling between system and bath, and \emph{the
role of the decoupling procedure is to minimize this distance}.

Using the triangle inequality on $\|\rho (T)-\rho ^{0}(T)+\rho ^{0}(T)-\rho
^{\mathrm{ideal}}(T)\|_{1}$ we have
\begin{equation}
D_{\mathrm{tot}}\leq D_{\mathrm{DD}}+D_{\mathrm{id}},  \label{eq:Del}
\end{equation}%
which shows that minimizing the total error can be done by separately
minimizing the open-system decoupling error and the closed-system control
error.

In Appendix \ref{app:mixed} (see also \cite{LZK:08} for a more general
treatment) we prove the following lemma:

\begin{mylemma}
  \label{lemma:mixed}
  Let $U=\exp (-iA)$ where $A$ is hermitian. Then for any
submultiplicative norm
\begin{eqnarray}
\|UBU^{\dag }-B\| &\leq& \|B\|\min [2,e^{2\|A\|_{\infty }}-1] \\
&\overset{2\|A\|_{\infty }\leq 1}{\leq }& 2\|B\|\min [1,(e-1)\|A\|_{\infty }]
\notag
\end{eqnarray}
\end{mylemma}

By identifying $A$ with the error phase $\Phi _{E}(T)$ and $B$ with $\tilde{%
\rho}_{0}(0)$ this allows us to write%
\begin{eqnarray}
D_{\mathrm{DD}} &=&\frac{1}{2}\|U_{\mathrm{err}}(T,0)\tilde{\rho}(0)U_{%
\mathrm{err}}^{\dag }(T,0)-\tilde{\rho}_{0}(0)\|_{1}  \notag \\
&\leq &\frac{1}{2}\|\rho _{0}(0)\|_{1}\min [2,(e^{2\|\Phi _{E}(T)\|_{\infty
}}-1)]  \notag \\
&=&\min [1,\frac{1}{2}(e^{2\|\Phi _{E}(T)\|_{\infty }}-1)]  \notag \\
&\overset{2\|\Phi _{E}(T)\|_{\infty }\leq 1}{\leq }& 2\|\Phi
_{E}(T)\|_{\infty }.  \label{eq:DelDD}
\end{eqnarray}

Inequality (\ref{eq:DelDD}) shows that minimization of the error phase $\Phi
_{E}(T)$ is sufficient for minimizing the decoherence error $D_{\mathrm{DD}}$%
. Combining our bounds {[}Eqs. (\ref{eq:Del-id}), (\ref{eq:Del}), and (\ref%
{eq:DelDD})] we have the quantum fidelity lower bound between the actual and
ideal system state:%
\begin{eqnarray}
&&F_{Q}[\rho _{S}(T),\rho _{S}^{\mathrm{ideal}}(T)]  \label{eq:F_Q} \\
&\geq &1-D[\rho _{S}^{0}(T),\rho _{S}^{\mathrm{ideal}}(T)]-\min [1,\frac{1}{2%
}(e^{2\|\Phi _{E}(T)\|_{\infty }}-1)].  \notag
\end{eqnarray}%
In other words, minimization of the pure-system control distance together
with minimization of the error phase $\Phi _{E}(T)$ is sufficient for
minimization of the total distance $D_{S}$. Note that the bound we have
derived is not necessarily tight:\ it is possible to minimize $D_{\mathrm{DD}%
}$ and $D_{\mathrm{id}}$ simultaneously, rather than separately, as is done
in fault tolerant quantum error correction \cite{Preskill:99}.

\section{Error Estimates for Dynamically Decoupled Logic Gates}

\label{sec:error}

Our goal in this section is to estimate the error associated with the
decoupled evolution (i.e., the evolution in the presence of a DD\ pulse
sequence), relative to the decoherence-free evolution (no system-bath
coupling). The decoupled evolution at the end of a DD\ cycle is described by
the propagator $U_{\text{err}}(T,0)$ of Eq.~(\ref{eq:Uerr2}). The
appropriate dimensionless error parameter is the norm of the total error
phase $\|\Phi_{E}\|_\infty $ [Eq.~(\ref{eq:Phi_E})]. Our strategy for
estimating $\Phi_{E}$ will be to calculate approximations to the final
effective Hamiltonian and then to bound its norm.
As we showed in
Section \ref{sec:bounds}, in the limit that $\Phi_{E}$ vanishes the final
state is free of decoherence errors.

Our main technical tool is the following lemma. For a proof see \cite{LZK:08}%
.
\begin{mylemma}
\label{lemma1}Consider a quantum evolution generated by a time-dependent
Hamiltonian $H(s)=H_{0}(s)+V(s)$, $0\leq s\leq t$, with propagators
satisfying $dU(s,0)/ds=-iH(s)U(s,0)$ and
$dU_{0}(s,0)=-iH_{0}(s)U_{0}(s,0)$. Then there exists a time-dependent Hamiltonian $H_{\text{\textrm{eff}}}(t)$ such that
\begin{equation}
\exp[-itH_{\text{\textrm{eff}}}(t)]\equiv U_{0}^{\dag}(t,0)U(t,0).
\end{equation}
and the following inequality holds for any unitarily invariant norm:%
\begin{eqnarray}
\Vert H_{\text{\textrm{eff}}}\Vert & \leq & \frac{1}{t}\int_{0}^{t}ds\Vert
V\left(s\right)\Vert\equiv\langle\Vert V\Vert\rangle_{t}  \label{eq:ineq1} \\
& \leq & \sup_{0<s<t}\Vert V(s)\Vert.
\end{eqnarray}
\end{mylemma}

This lemma allows us to relate the strength of the effective interaction
picture Hamiltonian at the end of the evolution $H_{\text{\textrm{eff}}}(t)$
to the strength of the (time-dependent) perturbation $V$.

As a first application, let us relate $\|H_{\text{err}}^{P_{i}}\|_\infty$ to $%
\|H_{\text{err}}\|_\infty$. Comparing lemma \ref{lemma1} with Eq.~(\ref{eq:H_err^Pi}%
) and identifying $H_{0}$ with $H_{P}^{(i)}$ {[}and thus $U_{0}(t,0)$ with
the ideal pulse $P_{i}$], $V(t)$ with Ad$_{tH_{\text{sec}}}(H_{\text{err}})$
{[}and thus $H(t)$ with $H_{P}^{(i)}+$Ad$_{tH_{\text{sec}}}(H_{\text{err}})$%
], and $\delta H_{\text{eff}}(\delta)$ with $\delta H_{\text{err}%
}^{P_{i}}(\delta)$, we have%
\begin{equation}
\|H_{\text{err}}^{P_{i}}(\delta)\|_\infty\leq\langle\|H_{\text{err}%
}\|_\infty\rangle_{\delta}\leq\sup_{t_{i+1}-\delta<t<t_{i+1}}\Vert H_{\text{err}%
}(t)\Vert_\infty.  \label{eq:pulseerr}
\end{equation}
This means that the application of a pulse, with inclusion of the
system-bath coupling during the pulse as in Eq.~(\ref{eq:H_err^Pi}), does
not cause a growth in the error rate. This is rather remarkable and can be
summarized as ``pulses can't hurt\textquotedblright .\footnote{
Of course by other measures pulses \emph{can} hurt. For example, broadband
pulses can cause unwanted transitions. But for the purposes of our error
estimates the absence of growth of the error norm is the crucial aspect.}

Similarly, by setting $H_{0}=0$ and $V(t)=$Ad$_{tH_{\text{sec}}}(H_{\text{err%
}})$ in lemma \ref{lemma1}, we obtain for the free evolution:
\begin{equation}
\|H_{\text{err}}^{(i)}\|_\infty\leq\langle\|H_{\text{err}}\|_\infty\rangle_{\tau}\leq%
\sup_{t_{i}<t<t_{i}+\tau}\Vert H_{\text{err}}(t)\Vert_\infty.
\end{equation}

Now let us return to $H_{\text{m}}$ {[}Eq.~(\ref{eq:Hm})], i.e., the
Hamiltonian describing the total evolution over a DD\ cycle. From now on we
simply denote $\langle\|H_{\text{err}}\|_\infty\rangle_{\delta}$ and $\langle\|H_{%
\text{err}}\|_\infty\rangle_{\tau}$ by $\|H_{\text{err}}\|_\infty$. Then the last two
inequalities yield $\|D_{i}H_{\text{err}}^{P_{i}}D_{i}^{\dagger}\|_\infty\leq\|H_{%
\text{err}}\|_\infty$ and $\|D_{i}H_{\text{err}}^{(i)}D_{i}^{\dagger}\|_\infty\leq\|H_{%
\text{err}}\|_\infty$, so that%
\begin{equation}
\left\Vert H_{\mathrm{m}}(t)\right\Vert_\infty \leq\|H_{\text{err}}\|_\infty.
\label{eq:normHm}
\end{equation}

At this point we are ready to use the Magnus expansion to estimate the error
phase $\Phi _{E}(T)$. Recalling Eq.~(\ref{eq:Uerr-final}), the Magnus
expansion for the error phase is given by%
\begin{eqnarray}
\Phi _{E}(T) &=& \int_{0}^{T}ds_{1}H_{\mathrm{m}}(s_{1})
\label{eq:finmagnus} \\
&+& \frac{1}{2}\int_{0}^{T}ds_{1}\int_{0}^{s_{1}}ds_{2}[H_{\mathrm{m}%
}(s_{2}),H_{\mathrm{m}}(s_{1})]+\cdots ,  \notag
\end{eqnarray}
and converges as long as $\int_{0}^{T}ds_{1}\|H_{\mathrm{m}%
}(s_{1})\|_{\infty }<\pi $ {[}Eq.~(\ref{eq:Mag-conv})], i.e., a sufficient
condition for convergence is
\begin{equation}
T\|H_{\text{err}}\|_{\infty }<\pi .  \label{eq:convcond}
\end{equation}

We assume that our decoupling sequence $\{P_{i}\}$ is designed for
cancelling error terms up to the first order in the Magnus expansion, as in
Section \ref{sec:DD-back}. Accordingly we rewrite $\Phi_{E}$ as a sum of the
first order terms and a second order correction, which we can in principle
improve upon by designing a pulse sequence that cancels error terms up to a
higher order:
\begin{equation}
\Phi_{E}(T)=\int_{0}^{T}H_{\mathrm{m}}(s)ds+\Phi_{\text{2nd}}(T).
\label{eq:1stphase}
\end{equation}
In Appendix \ref{app:trunc} we prove the following lemma

\begin{mylemma}
  \label{lem:Mag-partial}
Consider a time-dependent Hamiltonian $H(t)$, $0\leq
t\leq T$, and the partial sum of $k$th and higher order terms in the
corresponding Magnus expansion:%
\begin{equation}
\Phi _{k}=\sum_{i=k}^{\infty }\Omega _{i}.
\end{equation}%
Assume the Magnus expansion converges. Then%
\begin{equation}
\Vert \Phi _{k}\Vert _{\infty }\leq c_{k}(T\sup_{0<t<T}\Vert H(t)\Vert
_{\infty })^{k}
\end{equation}%
where $c_{k}=O(1)$ is a constant.
\end{mylemma}

Thus, subject to Eq.~(\ref{eq:convcond}),
\begin{equation}
\|\Phi _{\text{2nd}}(T)\|_{\infty }\leq c(T\|H_{\text{err}}\|_{\infty })^{2}
\label{eq:normC}
\end{equation}%
for some constant $c$. The fact that starting from arbitrary $k$th order the
error phase is upper bounded by $(T\|H_{\text{err}}\|_{\infty })^{k}$ means
that the design of higher order pulse sequences can be very advantageous in
achieving improved convergence of the DD\ procedure (see also \cite%
{sengupta:037202,pryadko:085321,Uhrig:07}), but we will not pursue this here.

To calculate the first order integral in Eq.(\ref{eq:1stphase}) we separate
the pulse and free parts:%
\begin{eqnarray}
\int_{0}^{T}H_{\mathrm{m}}(s)ds &=&\Phi _{\text{pulse}}+\Phi _{\text{free}},
\notag \\
\Phi _{\text{pulse}} &\equiv &\sum_{i=1}^{N}\int_{t_{i}-\delta }^{t_{i}}H_{%
\mathrm{m}}(s)ds,  \notag \\
\Phi _{\text{free}} &\equiv &\sum_{i=0}^{N-1}\int_{t_{i}}^{t_{i}+\tau }H_{%
\mathrm{m}}(s)ds.
\end{eqnarray}%
Using Eq.(\ref{eq:normHm}), the error phase due to the pulses $\Phi _{\text{%
pulse}}$ is bounded by%
\begin{equation}
\|\Phi _{\text{pulse}}\|_{\infty }\leq \Delta \|H_{\text{err}}\|_{\infty },
\end{equation}%
where
\begin{equation}
\Delta \equiv N\delta
\end{equation}%
is the total length of the pulse durations. Without use of additional
techniques such as pulse shaping \cite{sengupta:037202,pryadko:085321} or
composite pulse sequences \cite{Brown:04}, the only means at our disposal to
minimize this error is to make the pulse duration $\delta $ small. Taking
stock, we have, so far:%
\begin{eqnarray}
\|\Phi _{E}(T)\|_{\infty } &\leq &\|\int_{0}^{T}H_{\mathrm{m}%
}(s)ds\|_{\infty }+\|\Phi _{\text{2nd}}(T)\|_{\infty } \\
&\leq &\Delta \|H_{\text{err}}\|_{\infty }+\|\Phi _{\text{free}}\|_{\infty
}+c(T\|H_{\text{err}}\|_{\infty })^{2}.  \notag
\end{eqnarray}

The {}\textquotedblleft free error\textquotedblright\ $\|\Phi _{\text{free}%
}\|_{\infty }$ is the target of the DD pulses. Explicitly, we have, using
Eq.~(\ref{eq:Hm}):%
\begin{equation}
\Phi _{\text{free}}=\tau \sum_{i=1}^{N}D_{i}H_{\text{err}}^{(i)}D_{i}^{%
\dagger }.  \label{eq:Phifree}
\end{equation}

The effective error Hamiltonians $H_{\text{err}}^{(i)}$ can be
Magnus-expanded to first order in $\tilde{H}_{\text{err}}(t)=$Ad$_{tH_{\text{%
sec}}}[H_{\text{err}}]$ {[}recall Eq.~(\ref{eq:Herrtilde})], so that the
higher order commutators arising from the time-ordering in the definition of
$H_{\text{err}}^{(i)}$, are included in a new term $C$ that will be absorbed
into $\Phi _{\text{2nd}}$. First,
\begin{eqnarray}
\exp [-i\tau H_{\text{err}}^{(i)}] &=&\mathcal{T}\exp
[-i\int_{t_{i-1}}^{t_{i-1}+\tau }\tilde{H}_{\text{err}}(t)dt]  \notag \\
&=&\exp [-i\int_{t_{i-1}}^{t_{i-1}+\tau }\tilde{H}_{\text{err}}(t)dt+C^{(i)}]
\notag \\
&=&\exp [-i\int_{t_{i-1}}^{t_{i-1}+\tau }\sum_{n=0}^{\infty }\frac{%
(-it)^{n}}{n!}\times  \notag \\
&&[_{n}H_{\text{sec}},H_{\text{err}}]dt+C^{(i)}],
\end{eqnarray}%
where by Lemma \ref{lem:Mag-partial}
\begin{eqnarray}
\|C^{(i)}\|_{\infty } &\leq &d(\tau \sup_{t_{i-1}\leq t\leq t_{i-1}+\tau }\|%
\tilde{H}_{\text{err}}(t)\|_{\infty })^{2}  \notag \\
&=&d(\tau \|H_{\text{err}}\|_{\infty })^{2},
\end{eqnarray}%
and where $d$ is a constant. The integrals yield%
\begin{eqnarray}
f_{n,j} &\equiv &\int_{t_{j-1}}^{t_{j-1}+\tau }\frac{(-it)^{n}}{n!}dt  \notag
\\
&=&\frac{(-i)^{n}}{(n+1)!}\left[ (t_{j-1}+\tau )^{n+1}-t_{j-1}^{n+1}\right]
\end{eqnarray}

Therefore we can define $H_{\text{err}}^{(i)}$ via the expansion
\begin{equation}
\tau H_{\text{err}}^{(i)}=\tau H_{\text{err}}+\sum_{n=1}^{%
\infty}f_{n,i}[_{n}H_{\text{sec}},H_{\text{err}}]+C^{(i)}.  \label{eq:sumexp}
\end{equation}

Returning now to $\Phi _{\text{free}}$ {[}Eq.~(\ref{eq:Phifree})], notice
that the first term ($H_{\text{err}}$) in Eq.(\ref{eq:sumexp}) does not
depend on $t_{i}$ and is singled out, so that we may write:%
\begin{eqnarray}
\Phi _{\text{free}} &=&\Phi _{\text{dec}}+\Phi _{\text{undec}}+C,  \notag \\
\Phi _{\text{dec}} &\equiv &\tau \sum_{i=1}^{N}D_{i}H_{\text{err}%
}D_{i}^{\dagger },  \notag \\
\Phi _{\text{undec}} &\equiv &\sum_{i=1}^{N}\sum_{n=1}^{\infty
}f_{n,i}[_{n}H_{\text{sec}},D_{i}H_{\text{err}}D_{i}^{\dagger }]  \notag \\
C &\equiv &\sum_{i=1}^{N}C^{(i)},  \label{eq:dec+undec}
\end{eqnarray}%
where%
\begin{eqnarray}
\left\Vert C\right\Vert _{\infty } &\leq &Nd(\tau \|H_{\text{err}}\|_{\infty
})^{2}  \notag \\
&=&\frac{1}{N}[d(T-\Delta )\|H_{\text{err}}\|_{\infty })^{2}].
\end{eqnarray}%
The purpose of the DD\ procedure is, of course, to cancel $\Phi _{\text{dec}%
} $ {[}recall Eq.~(\ref{eq:DDcond})]. Pulse sequences that cancel higher
order terms ($n\geq 1$ in $\Phi _{\text{undec}}$) can be found, but this
will not be pursued here. Thus in our case the undecoupled terms will be
given by $\Phi _{\text{undec}}$. Define
\begin{equation}
\beta \equiv \|H_{\text{sec}}\|_{\infty },\quad J\equiv \|H_{\text{err}%
}\|_{\infty }.  \label{eq:bJ}
\end{equation}%
Let us first note that, using the triangle inequality on $\Phi _{\text{undec}%
}=\Phi _{\text{free}}-\Phi _{\text{dec}}$ and Eqs. (\ref{eq:Phifree}),(\ref%
{eq:sumexp}):
\begin{eqnarray}
\left\Vert \Phi _{\text{undec}}\right\Vert _{\infty } &\leq &\left\Vert \Phi
_{\text{free}}\right\Vert _{\infty }  \notag \\
&\leq &\tau \sum_{i=1}^{N}\|D_{i}H_{\text{err}}^{(i)}D_{i}^{\dagger
}\|_{\infty }  \notag \\
&=&\tau \sum_{i=1}^{N}\|H_{\text{err}}^{(i)}\|_{\infty }\leq T\left\Vert H_{%
\text{err}}\right\Vert _{\infty }  \notag \\
&=&JT.  \label{eq:trivcon}
\end{eqnarray}%
This trivial upper bound simply means that the undecoupled error is bounded
above by {}{}\textquotedblleft do nothing\textquotedblright . Another upper
bound for $\Phi _{\text{undec}}$ can be found by making use of norm
submultiplicativity:%
\begin{equation}
\left\Vert \lbrack A,B]\right\Vert _{\infty }\leq 2\left\Vert A\right\Vert
_{\infty }\left\Vert B\right\Vert _{\infty },
\end{equation}%
for any pair of operators $A$ and $B$ in the combined system-bath Hilbert
space. Then:%
\begin{eqnarray}
\|\Phi _{\text{undec}}\|_{\infty } &\leq &\sum_{i=1}^{N}\sum_{n=1}^{\infty
}\left\vert f_{n,i}\right\vert \,\|[_{n}H_{\text{sec}},D_{i}H_{\text{err}%
}D_{i}^{\dagger }]\|_{\infty }  \notag \\
&\leq &\sum_{i=1}^{N}\sum_{n=1}^{\infty }\left\vert f_{n,i}\right\vert
(2\beta )^{n}J  \notag \\
&=&J\sum_{i=1}^{N}\sum_{n=1}^{\infty }\frac{(t_{i-1}+\tau
)^{n+1}-t_{i-1}^{n+1}}{(n+1)!}(2\beta )^{n}  \notag \\
&\leq &J\sum_{i=1}^{N}\sum_{n=1}^{\infty }\frac{(\tau +\delta
+t_{i-1})^{n+1}-t_{i-1}^{n+1}}{(n+1)!}(2\beta )^{n}  \notag \\
&=&J\sum_{n=1}^{\infty }\frac{(2\beta )^{n}}{(n+1)!}%
\sum_{i=1}^{N}t_{i}^{n+1}-t_{i-1}^{n+1}  \notag \\
&=&\frac{J}{2\beta }\sum_{n=1}^{\infty }\frac{T^{n+1}}{(n+1)!}(2\beta )^{n+1}
\notag \\
&=&JT\frac{(e^{2\beta T}-1-2\beta T)}{2\beta T},  \label{eq:weakcon}
\end{eqnarray}%
where in the penultimate equality we used $T=t_{N}$ and $t_{0}=0$.

Combining the bounds in Eqs. (\ref{eq:trivcon}),(\ref{eq:weakcon}) we obtain
the following bound on the strength of the undecoupled terms:%
\begin{equation}
\|\Phi _{\text{undec}}\|_{\infty }\leq JT\min \left[ 1,\frac{\exp (2\beta
T)-1}{2\beta T}-1\right] .
\end{equation}%
Combining the expressions for various parts of the total error phase $\Phi
_{E},$we obtain the following upper bound:

\begin{eqnarray}
\left\Vert \Phi _{E}\right\Vert _{\infty } &\leq &\|\Phi _{\text{2nd}%
}\|_{\infty }+\|\Phi _{\text{pulse}}\|_{\infty }+\|\Phi _{\text{dec}%
}\|_{\infty }  \notag \\
&&+\|\Phi _{\text{undec}}\|_{\infty }+\|C\|_{\infty }  \notag \\
&\leq &c(JT)^{2}+J\Delta +0  \notag \\
&&+JT\min \left[ 1,\frac{\exp (2\beta T)-1}{2\beta T}-1\right]  \notag \\
&&+\frac{1}{N}[dJ(T-\Delta ))^{2}].
\end{eqnarray}%
Now note that for fixed $N$ and $\Delta $ we can always write $d(T-\Delta )/%
\sqrt{N}$ as $c^{\prime }T$, where $c^{\prime }$ accounts for the shift and
rescaling of $T$. This allows us to absorb $\frac{1}{N}[d(T-\Delta )J)^{2}]$
into $c(TJ)^{2}$ (redefining $c$ in the process), so that:%
\begin{equation}
\left\Vert \Phi _{E}\right\Vert _{\infty }\leq c(JT)^{2}+J\Delta +JT\min
\left[ 1,\frac{\exp (2\beta T)-1}{2\beta T}-1\right] .  \label{eq:phieb}
\end{equation}%
This, in conjunction with Eq.~(\ref{eq:F_Q}), finally gives us the desired lower
bound on the quantum fidelity of one period of DD.

\section{Periodic Dynamical Decoupling}

\label{sec:long}

Note that in principle the bound (\ref{eq:phieb}) is appropriate for any DD
sequence, since the time $T$ is arbitrary (subject to the convergence of the
Magnus expansion) and the decoupling group can have arbitrarily many
elements. However, in practice DD pulse sequences have some deterministic
structure, such as periodicity or self-similarity, or are random. Structure
generally results in improved performance under appropriate
circumstances \cite{KL1,KL2,Yao:07,Witzel:07,Lee:07,Zhang:08,Uhrig:07},
and hence the bound (\ref{eq:phieb}) may be too weak.

In this section we apply the idea of encoded operations and dynamical
decoupling to the periodic case (PDD) \cite{Viola:99} and derive the final-time error bound.
The encoded operation consists of the switching of a physical Hamiltonian
corresponding to a logical Hamiltonian for a duration of $T_m$.
This switching period $T_m$ is punctuated at various points by
the action of dynamical decoupling operations. In the previous section, the
analysis was performed for a basic cycle of $N$ pulses. In this
section we consider what happens when this sequence is applied $m$
times.

Consider a basic decoupling sequence $p$ designed to cancel all terms in $%
\tilde{H}_{\text{err}}$, as in Eq.~(\ref{eq:genericseq}):
\begin{equation}
p\left(\mathtt{f}_{\tau}\right)=P_{N}\mathtt{f}_{\tau}P_{N-1}\mathtt{f}%
_{\tau}P_{N-2}\cdots P_{1}\mathtt{f}_{\tau},
\end{equation}
where $\{P_{i}\}$ is the sequence of $N$ decoupling pulses and $\mathtt{f}%
_{\tau}$ denotes a {}{}``pause\textquotedblright\ of duration $\tau=\frac{T}{%
N}$ in decoupling, during which the control Hamiltonian $H_{\text{ctrl}}(t)$
generating the encoded logic gate is operative. Consider now the longer
\emph{periodic sequence} $\text{PDD}_{m}$ formed by repeating $p\left(%
\mathtt{f}_{\tau}\right)$ $m$ times to obtain a sequence of length $T_{\text{%
long}}=mT$ with $N_{m}=mN$ pulses:%
\begin{equation}
\text{PDD}_{m}=\prod_{j=1}^{m}p\left(\mathtt{f}_{\tau}\right).
\end{equation}
In the absence of encoded operations the sequence $p\left(\mathtt{f}%
_{\tau}\right)$ is designed to cancel dynamics up to the first order. The
longer sequence $\text{PDD}_{m}$ has the same canceling properties as the
sequence $p$ in the limit of $\tau\rightarrow0$.

So far we have not been specific about how we implement the encoded
operation. Namely, we have considered general time-dependent control
Hamiltonians. For simplicity, from now on we consider the following simple
method for realizing encoded operations. First, we only implement one logic
gate during each PDD sequence. In other words, a new logic gate requires a
new PDD\ sequence. Second, each logic gate is implemented in terms of a
constant control Hamiltonian. Thus, if ideally we wish to implement $U_{%
\text{ctrl}}(T_m)=\exp(-iT_mH_{\text{ctrl}%
})=\exp(-i\theta R)$ {[}Eq.~(\ref{eq:encodedR})], where $H_{\text{ctrl}%
}=\lambda R$ with $\lambda$ the magnitude of $H_{\text{ctrl}}$ and $%
\theta=\lambda T_m$ the phase, then in practice we will
implement the decoupling-free intervals as
\begin{equation}
\mathtt{f}_{\tau}=\exp\left[-i\tau(H_{\text{err}}+H_{B})-i\frac{\theta}{N_{%
m}}R\right].
\end{equation}
I.e., the encoded operation is implemented little-by-little, using $N_{\text{%
long}}$ equal $N_{m}$th root segments.

Let us now find a bound on the fidelity of PDD$_{m}$ in this setting. Since
we implement the encoded operations using the fixed step $\mathtt{f}_{\tau}$%
, the propagator for each cycle in the periodic sequence is the same, and
hence so is the error phase at the end of each DD\ cycle. Formally, the
total propagator in the interaction picture is simply {[}recall Eq.~(\ref%
{eq:Uerr-final})]:

\begin{eqnarray}
U_{\text{err}}(T_m,0) &=&U_{\text{err}}(T_m,(m-1)T)
\notag \\
&&\cdots U_{\text{err}}(2T,T)U_{\text{err}}(T,0)  \notag \\
&=&\prod\limits_{j=1}^{m}e^{-i\Phi _{E}(jT)}=[e^{-i\Phi _{E}(T)}]^{m}  \notag
\\
&=&e^{-im\Phi _{E}(T)}\equiv e^{-i\Phi _{\text{PDD}_{m}}}.
\end{eqnarray}

Recalling our fidelity bound Eq.~(\ref{eq:F_Q}), our task is to estimate the
norm of the error phase associated with the periodic sequence after time $T_{%
m}$, i.e., $\Phi _{\text{PDD}_{m}}\equiv m\Phi _{E}(T)$. It thus
follows immediately from Eq.~(\ref{eq:phieb}) that:
\begin{eqnarray}
\|\Phi _{\text{PDD}_{m}}\|_{\infty } &\leq &c(JT_m)^{2}/m+N_{%
m}J\delta  \label{eq:PDD-final} \\
&&+JT_m\min \left[ 1,\frac{\exp (2\beta T_m/m)-1}{%
2\beta (T_m/m)}-1\right] .  \notag
\end{eqnarray}%
In the limit of $\delta =0$ and $\beta T_m\ll 1$, we have
(second order Taylor expansion):
\begin{equation}
\|\Phi _{\text{PDD}_{m}}\|_{\infty }\leq m(cJ^{2}+J\beta )T^{2}.
\label{eq:PDD-approx}
\end{equation}

We postpone an analysis of this result until Section~\ref{sec:conc}.

\section{Example:\ Quantum Computation using the Heisenberg Interaction}

\label{sec:Heis-DD}

The commutation condition (\ref{eq:commcond}) is crucial to our results. At
first sight it appears that one cannot satisfy it while having non-trivial
decoupling operations. However, as pointed out in Ref. \cite{Lidar:AQC-DD},
it can be satisfied using the double commutant construction, which we now
explain.

The decoupling group $\mathcal{G}$ induces a decomposition of the system
Hilbert space $\mathcal{H}_{S}$ via its group algebra $\mathbb{C}\mathcal{G}$
and its commutant $\mathbb{C}\mathcal{G}^{\prime }$, as follows \cite%
{Viola:00a,Zanardi:99d}:%
\begin{align}
\mathcal{H}_{S}& \cong \bigoplus_{J}\mathbb{C}^{n_{J}}\otimes \mathbb{C}%
^{d_{J}},  \label{eq:Hsplit} \\
\mathbb{C}\mathcal{G}& \cong \bigoplus_{J}I_{n_{J}}\otimes M_{d_{J}},\quad
\mathbb{C}\mathcal{G}^{\prime }\cong \bigoplus_{J}M_{n_{J}}\otimes I_{d_{J}}.
\end{align}%
Here $n_{J}$ and $d_{J}$ are, respectively, the multiplicity and dimension
of the $J$th irrep of the unitary representation chosen for $\mathcal{G}$,
while $I_{N}$ and $M_{N}$ are, respectively, the $N\times N$ identity matrix
and unspecified complex-valued $N\times N$ matrices. We encode the
computational state into (one of) the left factors $\emph{C}_{J}\equiv
\mathbb{C}^{n_{J}}$, i.e., each such factor (with $J$ fixed) represents an $%
n_{J}$-dimensional code $\emph{C}_{J}$ storing $\log _{d}n_{J}$ qu$d$its.
Our DD\ pulses act on the right factors. As shown in \cite{Viola:00a}, the
dynamically decoupled evolution on each factor (code) $\emph{C}_{J}$ will be
noiseless in the ideal limit $w,\tau \rightarrow 0$ iff $\Pi _{\mathcal{G}%
}(S_{\alpha })=\bigoplus_{J}\lambda _{J,\alpha }I_{n_{J}}\otimes I_{d_{J}}$
[the projection $\Pi _{\mathcal{G}}$ was defined in Eq.~(\ref{eq:proj})] for
all system operators $S_{\alpha }$ in $H_{SB}$, whence $H_{\mathrm{eff}%
}^{(1)}=\bigoplus_{J}\left[ \left( I_{n_{J}}\otimes I_{d_{J}}\right) \right]
_{S}\otimes \left[ \sum_{\alpha }\lambda _{J,\alpha }B_{\alpha }\right] _{B}$%
. Thus, assuming the latter condition is met, \emph{under the action of
ideal DD the action of }$H_{\mathrm{eff}}^{(1)}$ \emph{on the code} $\emph{C}%
_{J}$\emph{\ is proportional to }$I_{n_{J}}$\emph{, i.e., is harmless}.
Quantum logic is enacted by the elements of $\mathbb{C}\mathcal{G}^{\prime }$%
. Dynamical decoupling operations are enacted via the elements of $\mathbb{C}%
\mathcal{G}$. \emph{We satisfy condition} (\ref{eq:commcond}) \emph{because }%
$[\mathbb{C}\mathcal{G},\mathbb{C}\mathcal{G}^{\prime }]=0$.

As an example, consider quantum computation\ with the Heisenberg interaction
\cite{Bacon:99a,Kempe:00,DiVincenzo:00a}. For the purposes of quantum
computing with electron spins in quantum dots, where a linear system-bath
interaction of the form
\begin{equation}
H_{SB}^{\mathrm{lin}}=\sum_{\alpha =x,y,z}\sum_{j}\sigma _{j}^{\alpha
}\otimes B_{j}^{\alpha },
\end{equation}
is the dominant source of decoherence due to hyperfine coupling to impurity
nuclear spins, it is convenient to use only Heisenberg interactions $H_{%
\mathrm{Heis}}=\sum_{i<j}J_{ij}\vec{\sigma}_{i}\cdot \vec{\sigma}_{j}$,
without physical-level single-qubit gates \cite{DiVincenzo:00a}. Here $\vec{%
\sigma}_{j}=(\sigma _{j}^{x},\sigma _{j}^{y},\sigma _{j}^{z})$ are the Pauli
matrices on the $j$th system qubit and $B_{j}^{\alpha }$ are arbitrary bath
operators. To beat $H_{SB}^{\mathrm{lin}}$ we use the Abelian
\textquotedblleft universal decoupling group\textquotedblright\ \cite%
{Viola:99} $\mathcal{G}_{\mathrm{uni}}=\{I,X,Y,Z\}$, where $%
X=\bigotimes_{j}\sigma _{j}^{x}$, $Y=\bigotimes_{j}\sigma _{j}^{y}$, $%
Z=\bigotimes_{j}\sigma _{j}^{z}$. It is simple to verify that $\Pi _{%
\mathcal{G}_{\mathrm{uni}}}(H_{SB}^{\mathrm{lin}})=0$. This is compatible
with using $\mathcal{G}_{\mathrm{uni}}$ to eliminate $H_{SB}^{\mathrm{lin}}$%
, \emph{since the global }$X,Y$ \emph{and }$Z$\emph{\ pulses commute with
the Heisenberg interaction}. I.e., this is an explicit example of Eq.~(\ref%
{eq:commcond}), where we identify $H_{\text{ctrl}}$ with $H_{\mathrm{Heis}}$%
, and $H_{\text{DD}}$ with the Hamiltonian generating the global pulses $X$,
$Y$ and $Z$, namely $\sum_{j}\sigma _{j}^{\alpha }$, $\alpha =x,y,z$. As is
well known \cite{Bacon:99a,Kempe:00,DiVincenzo:00a}, universal quantum
computation is possible using only the Heisenberg interaction provided
qubits are encoded into appropriate decoherence-free subspaces or subsystems.

\section{Discussion and Conclusions}

\label{sec:conc}

Let us now combine our two main results, Eqs. (\ref{eq:F_Q})\ and (\ref%
{eq:PDD-final}), for $m$ PDD\ cycles, each of duration $T$, i.e., of total
duration $T_m$, involving $N_{m}$ pulses each of width
$\delta $ and interval $\tau $:
\begin{widetext}
\begin{eqnarray}
F_{Q}[\rho _{S}(T_m),\rho _{S}^{\mathrm{ideal}}(T_m)]  &\geq&
1-D[\rho _{S}^{0}(T_m),\rho _{S}^{\mathrm{ideal}}(T_{%
m})]  -\min [1,\frac{1}{2}(e^{2\|\Phi _{\text{PDD}_{m}}\|_{\infty }}-1)], \\
\|\Phi _{\text{PDD}_{m}}\|_{\infty } &\leq& c(JT_m)^{2}/m+N_{m}J\delta
+JT_m\min \left[ 1,\frac{\exp (2\beta T_m/m)-1}{%
2\beta (T_m/m)}-1\right] ,  \label{eq:Phi-DD}
\end{eqnarray}%
\end{widetext}
or, in simplified form (assuming $\beta T_m\ll 1$, $\|\Phi _{%
\text{PDD}_{m}}\|_{\infty }\leq 1/2$, and zero-width pulses):

\begin{eqnarray}
&&F_{Q}[\rho _{S}(T_m),\rho _{S}^{\mathrm{ideal}}(T_{m%
})]  \label{eq:F-appr} \\
&\geq &1-D[\rho _{S}^{0}(T_m),\rho _{S}^{\mathrm{ideal}}(T_{%
m})]-2m(cJ^{2}+J\beta )T^{2}.  \notag
\end{eqnarray}%
We remind the reader that the term $D[\rho _{S}^{0}(T_m),\rho
_{S}^{\mathrm{ideal}}(T_m)]$ is the error due to control
imperfections in the uncoupled setting, and must be dealt with using methods
such as fault tolerant quantum error correction, composite pulses, or pulse
shaping.

The term $c(JT_m)^{2}/m=mcJ^{2}T^{2}$ in Eq.~(\ref{eq:Phi-DD})
is a bound on the error due to the fact that we have terminated the Magnus
expansion at second order. It can in principle be improved by performing a
more careful higher order perturbation theory analysis. The term $N_{\text{%
long}}J\delta $ is the error due to finite pulse width. This error can be
improved by using pulse shaping techniques \cite{Kofman:01,sengupta:037202}.
The last term in Eq.~(\ref{eq:Phi-DD}) is a bound on the undecoupled errors,
i.e., errors due to imperfect decoupling. Considering the zero-width pulse
limit, Eq.~(\ref{eq:F-appr}), we see that provided the number of cycles $m$ scales more slowly than $[2(cJ^{2}+J\beta )T^{2}]^{-1}$, i.e., if%
\begin{equation}
m=o\{[2(cJ^{2}+J\beta )T^{2}]^{-1}\},  \label{eq:m}
\end{equation}%
the fidelity is guaranteed to be dominated by the error $D[\rho _{S}^{0}(T_{%
m}),\rho _{S}^{\mathrm{ideal}}(T_m)]$ due to control
imperfections (the \textquotedblleft little-$o$\textquotedblright\ notation
means that the right-hand side dominates the left-hand side asymptotically).

We also recall that $\beta \equiv \|H_{\text{sec}}\|_{\infty }=\|H_{\text{%
ctrl}}\otimes I_{B}+I_{S}\otimes H_{B}\|_{\infty }\leq \|H_{\text{ctrl}%
}\|_{\infty }+\|H_{B}\|_{\infty }$ and $J\equiv \|H_{\text{err}}\|_{\infty
}=\|H_{SB}+H_{S,\mathrm{res}}\|_{\infty },$ where $H_{SB}$ is the
system-bath interaction Hamiltonian and $H_{S,\mathrm{res}}$ are residual
undesired pure-system terms that do not commute with $H_{\text{ctrl}}$.
Expressing the system-bath interaction as $H_{SB}=\sum_{\alpha }S_{\alpha
}\otimes B_{\alpha }$ (sum over system times bath operators), we have $J\leq
\sum_{\alpha }\|S_{\alpha }\|_{\infty }\|B_{\alpha }\|_{\infty }+\|H_{S,%
\mathrm{res}}\|_{\infty }$. For local Hamiltonians involving $n$ system
qubits we can reasonably expect $J\propto n$ (e.g, for electron spin qubits,
each of which is coupled to a local bath of nuclear spin impurities).
Similarly, we have $\|H_{\text{ctrl}}\|_{\infty }\propto n$ (assuming full
parallelism in the operation of the quantum computer). The norm of the
pure-bath Hamiltonian ($\|H_{B}\|_{\infty }$) may be very large, though in
practice it is always finite due to a high-energy cutoff or spatial cutoff
determining the relevant bath degrees of freedom. Assuming that we are
dealing with a bath for which $\|H_{B}\|_{\infty }\propto Mn$ (appropriate
spatial cutoff, such that the $n$ qubits couple to a bath with $M$ degrees
of freedom, where $M$ can be very large), we also have $\beta \propto n$.
Thus, we have from Eq.~(\ref{eq:m}) that for fixed $T$,
\begin{equation}
m\sim c^{\prime }n^{-2+\varepsilon },
\end{equation}%
where $c^{\prime }$ is a dimensionless constant involving the various energy
scales of the problem and $\varepsilon >0$. This last result establishes
that using PDD with fixed cycle time, there is a tradeoff between the number
of cycles and the size of the quantum register, i.e., there is a limit on
scalability. On the other hand, the complete inequality suggested by Eq.~(%
\ref{eq:m}) is%
\begin{equation}
\sqrt{m}T\ll \lbrack 2(cJ^{2}+J\beta )]^{-1/2}\sim n^{-1},
\end{equation}%
so that a better strategy might be to invest resources in shrinking the
cycle duration $T$ with $n$, so as to increase the number of cycles $m$.

Ultimately, based on various comparative studies
\cite{Viola:06,KL2,Witzel:07,Zhang:08}, we expect that there are
strategies that
will outperform PDD\ altogether and will lead to much improved scalability.
Such strategies are concatenated DD \cite{KL1,KL2,Yao:07,Witzel:07,Zhang:08}%
, randomized DD \cite{Viola:05,Viola:06,Santos:06}, and specially tailored
DD such as the sequence proposed in \cite{Uhrig:07} for the diagonal
spin-boson model. We expect that the rigorous analysis we have presented
here will prove useful in the analysis of these more elaborate pulse
sequences.

\acknowledgments
D.A.L. was sponsored by NSF under grants CCF-0523675 and CCF-0726439,
and by the United States Department of Defense. The views and
conclusions contained in this document are those of the authors and
should not be interpreted as representing the official policies,
either expressly or implied, of the U.S. Government.

\appendix
    {}
\section*{Appendix}
We provide background and prove the various Lemmas found in the main
text. For convenience we restate all the Lemmas in this appendix.

\section{Magnus expansion}

\label{app:Magnus}

This section is a brief summary of \cite{Iserles:02,Casas:07}. The Magnus
expansion is a method for solving first-order operator-valued linear
differential equations:%
\begin{eqnarray}
\frac{dU(t,0)}{dt} &=&-iH(t)U,\,\quad t\geq 0,\quad  \label{eq:Udot} \\
U(0) &=&I.  \notag
\end{eqnarray}%
Here $H(t)$ can be any bounded linear operator. When $H(t)$ is hermitian
(the only case we consider) Eq.~(\ref{eq:Udot}) is the time-dependent Schr%
\"{o}dinger equation and the Magnus expansion provides a unitary
perturbation theory, in contrast to the Dyson series. The unitary nature of
the Magnus expansion is one of it most appealing features.

The formal solution of Eq.~(\ref{eq:Udot}) is the time-ordered integral
\begin{eqnarray}
U(t) &=&\lim_{N\rightarrow \infty }\prod_{j=0}^{N}\exp \left[ -i\frac{t}{N}%
H\left( \frac{jt}{N}\right) \right]  \notag \\
&\equiv &\mathcal{T}\exp [-i\int_{0}^{t}H(s)ds].
\end{eqnarray}%
The Magnus expansion represent the solution in the form $U(t)=\exp [-i\Omega
(t)]$ and expresses $\Omega (t)$ in a series expansion. When $H(t)$ commutes
with $\int_{0}^{t}H(s)ds$ the solution is $U(t)=\exp [-i\int_{0}^{t}H(s)ds]$%
, $t\geq 0$ (no time-ordering). Otherwise the solution is an infinite
series:
\begin{equation}
U(t,0)=\lim_{n\rightarrow \infty }e^{iM_{n}(t)}
\end{equation}%
where $M_{n}(t)$ is the hermitian operator%
\begin{equation}
M_{n}=\sum_{i=1}^{n}\Omega _{i}  \label{eq:Mn}
\end{equation}%
where%
\begin{eqnarray}
&&\Omega _{i}[H(t)]_{0}^{t} \\
&&=\sum_{j}c_{j,i}\underset{0<t_{1}<\cdots <t_{n}<t}{\int_{t_{1}}%
\int_{t_{2}}\cdots \int_{t_{n}}}[[H(t_{1}),\cdots ,H(t_{n})]]dt_{n}\cdots
dt_{1} ,  \notag
\end{eqnarray}
where $[[H(t_{1}),\cdots ,H(t_{n})]]$ denotes an $n$th level nested
time-ordered commutator expression between $H(t_{i})$, and the coefficients $%
c_{j,i}$ are recursively defined and can be computed to any order. The first
few terms are:%
\begin{eqnarray}
\Omega _{1}& =&\int_{0}^{t}H(t_{1})dt_{1}  \label{eq:Magnus} \\
\Omega _{2}&=&\frac{1}{2}\int_{0}^{t}dt_{1}%
\int_{0}^{s}dt_{2}[H(t_{1}),H(t_{1})]  \notag \\
\Omega _{3}& =&\frac{1}{12}\int_{0}^{t}dt_{1}\int_{0}^{t_{1}}dt_{2}%
\int_{0}^{t_{2}}dt_{3}[H(t_{3}),[H(t_{2}),H(t_{1})]]  \notag \\
& +& \frac{1}{4}\int_{0}^{t}dt_{1}\int_{0}^{t_{1}}ds^{\prime
}\int_{0}^{t_{2}}dt_{3}[[H(t_{3}),H(t_{2})],H(t_{1})].  \notag
\end{eqnarray}
A sufficient (but not necessary) condition for absolute convergence of the
Magnus series $M_{n}(t)$ in the interval $[0,t)$ is \cite{Casas:07}:
\begin{equation}
\int_{0}^{t}\left\Vert H(s)\right\Vert _{\infty }ds<\pi .
\label{eq:Mag-conv}
\end{equation}

\section{Evolution lemma for a switched Hamiltonian}

\label{app:Udecomp}

We prove Lemma~\ref{lem1}:

The propagator generated by a ``switched Hamiltonian\textquotedblright
\begin{equation*}
H(t)=H_{i}(t)\quad t_{i-1}<t<t_{i},\quad i=1,...,N,
\end{equation*}
can be decomposed into corresponding segments:%
\begin{equation}
U(t_{N},t_{0})=U(t_{N},t_{N-1})\cdots U(t_{1},t_{0}),  \label{eq:decomp1}
\end{equation}
where $U(t_{i+1},t_{i})$, with $t_{i}\leq t\leq t_{i+1}$, satisfies
the Schr\"{o}dinger equation 
\begin{equation}
\frac{dU(t,t_{i})}{dt}=-iH(t)U(t,t_{i}),\quad U(t_{i},t_{i})=I.
\label{eq:Schr1}
\end{equation}

\begin{proof}
Denote the propagator generated by a time-dependent Hamiltonian $H(t)$
starting from an initial time $t_{0}$ by $U(t,t_{0})$. Evolving backward in
time from $t_{i}$ to $t_{0}$, followed by a forward in time evolution from $%
t_{0}$ to $t$ yields a net evolution from $t_{i}$ to $t$:
\begin{equation}
U(t,t_{i})=U(t,t_{0})U(t_{i},t_{0})^{\dag }.  \label{eq:bf}
\end{equation}%
Letting $t=t_{N}$ and $t_{i}=t_{N-1}$ we thus have:%
\begin{equation}
U(t_{N},t_{0})=U(t_{N},t_{N-1})U(t_{N-1},t_{0}).
\end{equation}%
Repeating this via $U(t_{N-1},t_{0})=$ $U(t_{N-1},t_{N-2})U(t_{N-2},t_{0})$
etc. we arrive at Eq.~(\ref{eq:decomp1}). To prove that $U(t,t_{i})$
satisfies Eq.~(\ref{eq:Schr1}) we differentiate Eq.~(\ref{eq:bf}) with
respect to $t$:%
\begin{eqnarray}
\frac{dU(t,t_{i})}{dt} &=&\frac{dU(t,t_{0})}{dt}U(t_{i},t_{0})^{\dag }
\notag \\
&=&-iH(t)U(t,t_{0})U(t_{i},t_{0})^{\dag }  \notag \\
&=&-iH(t)U(t,t_{i}).
\end{eqnarray}
\end{proof}

\section{Norms and Distances}

\label{app:norms}

Throughout this work we use unitarily invariant norms on bounded operators $%
A $ \cite{Bhatia:book}(Ch.4):
\begin{equation}
\|A\|=\|UAV\|\quad U,V\text{ unitary.}
\end{equation}
A norm is \emph{weakly} unitarily invariant if $\|A\|=\|UAU^{\dag}\|$ for
every unitary $U$. Obviously, if a norm is unitarily invariant then it is
also weakly unitarily invariant. In addition to being subadditive, i.e.,
satisfying the triangle inequality (by definition of a norm) $%
\|A+B\|\leq\|A\|+\|B\|$, unitarily invariant norms are also
submultiplicative \cite{Bhatia:book}(p.94):
\begin{equation}
\|AB\|\leq\|A\|\,\|B\|.
\end{equation}
Define
\begin{equation}
|A|\equiv\sqrt{A^{\dag}A}.
\end{equation}

The set of all square matrices, together with a submultiplicative norm, is
an example of a Banach algebra, and every $C^{\ast }$ algebra is a Banach
algebra. Note that not all matrix norms are submultiplicative. For example,
if we define $\left\Vert A\right\Vert _{\Delta }=\max_{ij}|a_{ij}|$ then for
the matrices $A=B=\left(
\begin{array}{cc}
1 & 1 \\
0 & 1%
\end{array}%
\right) $ we have $\left\Vert A\right\Vert _{\Delta }=\left\Vert
B\right\Vert _{\Delta }=1$ but $\left\Vert AB\right\Vert _{\Delta }=2$.

We now give three important examples \cite{Bhatia:book}(p.91-92).

The \emph{trace norm} is%
\begin{equation}
\|A\|_{1}\equiv \mathrm{tr}(|A|)\overset{A^{\dag }=A}{=}\mathrm{tr}A.
\end{equation}%
Note that if $\rho $ is a density matrix then $\|\rho \|_{1}=\mathrm{tr}\rho
=1$. The trace distance $D(\rho _{1},\rho _{2})\equiv \frac{1}{2}\|\rho
_{1}-\rho _{2}\|_{1} $, plays a special role since it captures the
measurable distance between different density matrices $\rho _{1}$ and $\rho
_{2}$ \cite{Aharonov:98a}. Namely, $D(\rho _{1},\rho _{2})$ is an achievable
upper bound on the trace distance between probability distributions arising
from measurements $P$ performed on $\rho _{1}$ and $\rho _{2}$ \cite%
{Nielsen:book}(Theorem 9.1), in the sense that $D(\rho _{1},\rho
_{2})=\max_{P}(\langle P\rangle _{1}-\langle P\rangle _{2})$, where $P\leq I$
is a positive operator, and $\langle P\rangle _{i}=\mathrm{tr}(P\rho _{i})$.

The \emph{Frobenius (or Hilbert-Schmidt) norm}

\begin{eqnarray}
\|A\|_{2} &\equiv &\sqrt{\langle A,A\rangle }=\sqrt{\mathrm{tr}(A^{\dag }A)}
\notag \\
&=&(\sum_{ij}|a_{ij}|^{2})^{1/2},
\end{eqnarray}%
(where $A$ has matrix elements $a_{ij}$) is the norm induced by the
Hilbert-Schmidt inner product
\begin{equation}
\langle A,B\rangle \equiv \mathrm{tr}A^{\dag }B.
\end{equation}

Finally, the \emph{operator norm} is $\|A\|_{\infty }\equiv
s_{1}(A)=\sup_{\|\psi \|=1}\|A|\psi \rangle \|$, where $s_{1}(A)$ is the
first (largest)\ singular value of $A$, i.e., the largest eigenvalue of $|A|$%
.

All three norms are multiplicative with respect to the tensor product \cite{Watrous:04}%
(Ch.2):%
\begin{equation}
\|A\otimes B\|_{i}=\|A\|_{i}\|B\|_{i}\quad i=1,2,\infty .  \label{eq:mult}
\end{equation}%
They satisfy the ordering%
\begin{equation}
\|A\|_{\infty }\leq \|A\|_{2}\leq \|A\|_{1}.  \label{eq:norm-ord}
\end{equation}%
Another useful inequality is \cite{Watrous:04}(Ch.2)
\begin{equation}
\|ABC\|\leq \|A\|_{\infty }\|B\|\|C\|_{\infty },
\end{equation}%
where $\left\Vert \cdot \right\Vert $ denotes any unitarily invariant norm.
A special case of this is obtained by setting $A$, $B$ or $C=I$:%
\begin{equation}
\|AB\|\leq \|A\|_{\infty }\|B\|,\|A\|_{\infty }\|B\|_{\infty },\|B\|_{\infty
}\|A\|.  \label{eq:inf}
\end{equation}%
An important inequality we need relates the norm of the partial trace
and the norm of the operator being traced over (for a proof see \cite{LZK:08}%
):%
\begin{eqnarray}
\|\mathrm{tr}_{B}X\|_{i} &\leq &d_{i}\|X\|_{i}\quad (i=1,2,\infty ),  \notag
\\
d_{1} &=&1,\quad  \notag \\
d_{2} &=&\sqrt{\dim (\mathcal{H}_{B})},\quad  \notag \\
d_{\infty } &=&\dim (\mathcal{H}_{B}),  \label{eq:part-trace}
\end{eqnarray}%
where $X$ is a linear operator over the tensor product Hilbert space $%
\mathcal{H}_{S}\otimes \mathcal{H}_{B}$. For the trace norm this is a
special case of the well known result that trace-preserving maps (in this
case the partial trace)\ are contractive \cite{Nielsen:book}.

\section{Norm to error phase inequality for mixed states}
\label{app:mixed}

We prove Lemma~\ref{lemma:mixed}:

Let $U=\exp (-iA)$ where $A$ is hermitian. Then for any
submultiplicative norm
\begin{eqnarray*}
\|UBU^{\dag }-B\| &\leq& \|B\|\min [2,e^{2\|A\|_{\infty }}-1] \\
&\overset{2\|A\|_{\infty }\leq 1}{\leq }& 2\|B\|\min [1,(e-1)\|A\|_{\infty }]
\end{eqnarray*}

\begin{proof}

First note that%
\begin{eqnarray}
e^{x}-1 &=&x\left( 1+\sum_{n=2}^{\infty }\frac{x^{n-1}}{n!}\right)  \notag \\
&\overset{x\leq 1}{\leq }& x\left( 1+\sum_{n=2}^{\infty }\frac{1}{n!}\right)
=(e-1)x.
\end{eqnarray}%
By a similar calculation we also get $\frac{e^{x}-1}{x}-1\overset{x\leq 1}{%
\leq }(e-2)x$.

By the triangle inequality%
\begin{equation}
\|UBU^{\dag }-B)\|\leq \|UBU^{\dag }\|+\|B\|\leq 2\|B\|.
\end{equation}%
On the other hand, using the Taylor expansion of $\exp (-i[A,\cdot ])$ we
have%
\begin{eqnarray}
\|UBU^{\dag }-B\| &=&\|\sum_{n=0}^{\infty }\frac{(-i)^{n}}{n!}[_{n}A,B]-B\|
\notag \\
&=&\|\sum_{n=1}^{\infty }\frac{i^{n}}{n!}[_{n}A,B]\|  \notag \\
&\leq &\sum_{n=1}^{\infty }\frac{1}{n!}\|[_{n}A,B]\|  \notag \\
&\leq &\sum_{n=1}^{\infty }\frac{2^{n}\|A\|_{\infty }^{n}}{n!}\|B\|  \notag
\\
&=&(e^{2\|A\|_{\infty }}-1)\|B\|  \notag \\
&\overset{2\|A\|_{\infty }\leq 1}{\leq }& 2(e-1)\|A\|\,_{\infty }\|B\|,
\end{eqnarray}%
where in the penultimate inequality we iterated
\begin{equation}
\left\Vert \lbrack A,B]\right\Vert \leq 2\left\Vert AB\right\Vert \leq
2\left\Vert A\right\Vert _{\infty }\left\Vert B\right\Vert ,
\end{equation}%
[where we used submultiplicativity together with Eq.~(\ref{eq:inf})] to get%
\begin{equation}
\|[_{n}A,B]\|\leq \|A\|\,_{\infty }\|[_{n-1}A,B]\|\leq \ldots \leq
\|A\|_{\infty }^{n}\|B\|.
\end{equation}
\end{proof}

\section{Magnus expansion truncation bound}
\label{app:trunc}

We prove Lemma~\ref{lem:Mag-partial}:

Consider a time-dependent Hamiltonian $H(t)$, $0\leq t\leq T$, and the
partial sum of $k$th and higher order terms in the corresponding Magnus
expansion:%
\begin{equation*}
\Phi _{k}=\sum_{i=k}^{\infty }\Omega _{i}.
\end{equation*}%
Assume the Magnus expansion converges in the trace norm. Then%
\begin{equation*}
\Vert \Phi _{k}\Vert \leq c_{k}(T\sup_{0<t<T}\Vert H(t)\Vert _{\infty })^{k}
\end{equation*}%
where $c_{k}=O(1)$ is a constant.

\begin{proof}
Define $h\equiv \sup_{0<t<T}\Vert H(t)\Vert _{\infty }$ and rescale $H(t)$
by $hT$:%
\begin{equation}
H^{\prime }(t)=\dfrac{H(t)}{hT}.
\end{equation}%
We can rewrite $\Omega _{i}$ as:%
\begin{eqnarray}
\Omega _{i}[H(t)]_{0}^{T} &=&(hT)^{i}\sum_{j}c_{j,i}\int_{0}^{T}\cdots
\int_{0}^{T}  \notag \\
&&[[H^{\prime }(t_{1}),\cdots ,H^{\prime }(t_{n})]]dt_{n}\cdots dt_{1}
\notag \\
&=&(hT)^{i}\Omega _{i}[H^{\prime }(t)]_{0}^{T}.  \label{eq:hTi}
\end{eqnarray}%
Recall the condition for absolute convergence of the Magnus expansion, Eq.~(%
\ref{eq:Mag-conv}). Since $\int_{0}^{T}\left\Vert H(t)\right\Vert dt\leq
T\sup_{0<t<T}\Vert H(t)\Vert $ a sufficient condition is:%
\begin{equation}
hT<1.  \label{eq:htl1}
\end{equation}%
Absolute convergence (convergence of the sum of absolute values) means that
if we define, for $k\geq 1$, the partial sum%
\begin{equation}
B_{n,k}\equiv \sum_{i=k}^{n}\Vert \Omega _{i}[H(t)]_{0}^{T}\Vert
\end{equation}%
then%
\begin{equation}
\lim_{n\rightarrow \infty }B_{n,k}=\beta _{k}[H(t)]_{0}^{T}<\infty ,
\end{equation}%
where $\beta _{k}$ is some functional of $H(t)$. Similarly for $H^{\prime
}(t)$:%
\begin{equation}
\lim_{n\rightarrow \infty }\sum_{i=k}^{n}\Vert \Omega _{i}[H^{\prime
}(t)]_{0}^{T}\Vert =\beta _{k}[H^{\prime }(t)]_{0}^{T}\equiv A_{k}=O(1).
\label{eq:absconv}
\end{equation}

Let us now focus on the partial sum of $k$th and higher order terms in the
Magnus expansion, $\Phi _{k}=\sum_{i=k}^{\infty }\Omega _{i}$. We can bound $%
\Phi _{k}$ in the following manner:%
\begin{eqnarray}
\Vert \Phi _{k}\Vert &\leq &\sum_{i=k}^{\infty }\Vert \Omega
_{i}[H(t)]_{0}^{T}\Vert  \notag \\
&\overset{\mathrm{Eq.(\ref{eq:hTi})}}{=}&(hT)^{k}\sum_{i=k}^{\infty
}(hT)^{i-k}\Vert \Omega _{i}[H^{\prime }(t)]_{0}^{T}\Vert  \notag \\
&\overset{\text{Eq.(\ref{eq:htl1})}}{\leq }&(hT)^{k}\sum_{i=k}^{\infty
}\Vert \Omega _{i}[H^{\prime }(t)]_{0}^{T}\Vert  \notag \\
&\overset{\text{Eq.(\ref{eq:absconv})}}{=}&(hT)^{k}A_{k} =O\left[ (hT)^{k}%
\right] .
\end{eqnarray}
\end{proof}


\end{document}